\documentclass[]{article}
\usepackage{proceed2e}
\usepackage{amsmath, amsthm, amssymb}
\usepackage{graphicx}
\usepackage[numbers]{natbib}
\usepackage[raggedright]{sidecap}
\usepackage[small]{caption}
\usepackage[centering,textheight=9.275in,textwidth=6.7395in,headsep=.75in]{geometry}
\usepackage{titleps}

\usepackage[activate={compatibility,compatibility}]{microtype}
 
 \newtheorem{thm}{Theorem}[section]

\newtheorem{lem}[thm]{Lemma}

 \newenvironment{comment}{
  \begin{list}{}{\rm
    \setlength{\topsep}{0pt}%
    \setlength{\leftmargin}{-0.2cm}%
    \setlength{\listparindent}{\parindent}%
    \setlength{\itemindent}{\parindent}%
    \setlength{\parsep}{\parskip}%
  }%
  \item[]}{\end{list}}

    \catcode`\@=11\relax	     %

\def\nofill {%
  \begingroup
    \obeyspaces
    \parskip=\z@
    \parindent=\z@
    \let\p@r=\par
    \def\par{\p@r \ifspacenext{\noindent}{}}%
    \obeylines
~~
}

\def\endnofill {%
    \p@r		%
~~
  \endgroup}

{\obeyspaces\global\let\sp@ce= \relax}

\def\ifspacenext #1#2{%
  \def\truet@ks{#1}%
  \def\falset@ks{#2}%
  \futurelet\next\ifsp@cenext}
\def\ifsp@cenext {%
  \ifx\next\sp@ce \truet@ks \else \falset@ks \fi}

\catcode`\@=12\relax	%

\title{Church: a language for generative models}

\author{ Noah D.~Goodman$^{*}$\!\!, Vikash K.~Mansinghka\thanks{\ The first two authors contributed equally to this work.},\\
{\bf Daniel Roy,  Keith Bonawitz \& Joshua B. Tenenbaum}\\  
MIT BCS/CSAIL \\
Cambridge, MA 02139 
} 

\newpagestyle{mystyle}{
  \sethead{\tiny \hspace{-.775in}In \textit{Proc.\ 24th Conf.\ Uncertainty in Artificial Intelligence (UAI)}, 220--229, 2008. This version contains minor corrections (May 31, 2008).}{}{}
  \setfoot{}{}{}
  }

\begin{document}
\thispagestyle{mystyle}
 
\maketitle 
 
\begin{abstract} 
Formal languages for probabilistic modeling enable re-use, modularity, and descriptive clarity, and can foster generic inference techniques.
We introduce \emph{Church}, a universal language for describing stochastic generative processes. Church is based on the Lisp model of lambda calculus, containing a pure Lisp as its deterministic subset. The semantics of Church is defined in terms of evaluation histories and conditional distributions on such histories. Church also includes a novel language construct, the stochastic memoizer, which enables simple description of many complex non-parametric models. We illustrate language features through several examples, including:
a generalized Bayes net in which parameters cluster over trials, infinite PCFGs, planning by inference, and various non-parametric clustering models.
Finally, we show how to implement query on any Church program, exactly and approximately, using Monte Carlo techniques.
\end{abstract}

\section{INTRODUCTION} 

Probabilistic models have proven to be an enormously useful tool in artificial intelligence, machine learning, and cognitive science. Most often these models are specified in a combination of natural and mathematical language, and inference for each new model is implemented by hand. Stochastic programming languages \cite[e.g.][]{Milch2005,Pfeffer2001,McAllester2008} aim to tame the model-building process by giving a formal language which provides simple, uniform, and re-usable descriptions of a wide class of models, and supports generic inference techniques.
In this paper we present the Church stochastic programming language (named for computation pioneer Alonzo Church),
a universal language for describing generative processes and conditional queries over them. Because this language is based on Church's lambda calculus, expressions, which represent generative models, may be arbitrarily composed and abstracted. 
The distinctive features of Church, and the main contributions of this paper, are: 1) a Lisp-like language specification in which we view \emph{evaluation as sampling} and \emph{query as conditional sampling}, 2) a \emph{stochastic memoizer}, which allows separate evaluations to share generative history and enables easy description of non-parametric probabilistic models, and, 3) generic schemes for exact and approximate inference, which implement the query primitive, so that any Church program may be run without writing special-purpose inference code.

\section{THE CHURCH LANGUAGE} 
  
The Church language is based upon a pure subset of the functional language Scheme \cite{r5rs}, a Lisp dialect. 
Church is a dynamically-typed, applicative-order language, in which procedures are first-class and expressions are values. 
Church expressions describe generative processes: the meaning of an expression is specified through a primitive procedure {\tt eval}, which samples from the process, and a primitive procedure {\tt query}, which generalizes {\tt eval} to sample conditionally. In true Lisp spirit, {\tt eval} and {\tt query} are ordinary procedures that may be nested within a Church program. Randomness is introduced through stochastic primitive functions; memoization allows random computations to be reused.

Church expressions have the form:  
 \begin{eqnarray*}
\text{expression} &::=& c\ |\ x\ |\ (e_{1}\ e_{2}\ ...)\ |\ (\text{\tt lambda}\ (x ...)\ e)\ | \\
& & (\text{\tt if}\ e_{1}\ e_{2}\ e_{3})\ |\ (\text{\tt define}\ x\ e)\ |\ (\text{\tt quote}\ e) \\
\end{eqnarray*}
Here $x$ stands for a variable (from a countable set of variable symbols), $e_{i}$ for expressions, and $c$ for a (primitive) constant. 
(We often write {\tt '$e$} as shorthand for {\tt (quote $e$)}.)

The constants include primitive data types (nil, Boolean, char, integer, fixed-precision real, etc.), and standard functions to build data structures (notably {\tt pair}, {\tt first}, and {\tt rest} for lists) and manipulate basic types (e.g. {\tt and}, {\tt not})\footnote{The primitive function {\tt gensym} deserves special note: {\tt (eval '(gensym) env)} returns a procedure ({\tt c, x, env}) where {\tt c} is a constant function which returns {\tt True} if {\tt x} is bound to the procedure ({\tt c, x, env}), and {\tt False} otherwise. Furthermore it is guaranteed that {\tt  (gensym (gensym))} evaluates to {\tt False} (i.e. each evaluation of {\tt gensym} results in a unique value).}.
As in most programming languages, all primitive types are countable; real numbers are approximated by either fixed- or floating-precision arithmetic.
A number of standard (deterministic) functions, such as the higher-order function {\tt map}, are provided as a standard library, automatically defined in the global environment. Other standard Scheme constructs are provided---such as {\tt (let ((a \emph{a-def}) (b \emph{b-def}) ...)~$body$)}, which introduces names that can be used in $body$, and is sugar for nested {\tt lambda}s.

Church \emph{values} include Church expressions, and procedures; if $v_{1} ... v_{n}$ are Church values the list $(v_{1} ... v_{n})$ is a Church value.
A Church \emph{environment} is a list of pairs consisting of a variable symbol and a value (the variable is \emph{bound} to the value); note that an environment is a Church value.
Procedures come in two types: \emph{Ordinary procedures} are triples, ({\tt body}, {\tt args}, {\tt env}), of a Church expression (the body), a list of variable symbols (the formal parameters, or arguments), and an environment. \emph{Elementary random procedures} are ordinary procedures that also have a \emph{distribution function}---a probability function that reports the probability $P( \text{ value } | \text{ {\tt env}, {\tt args}})$ of a return value from evaluating the body (via the {\tt eval} procedure described below) given {\tt env} and values of the formal parameters\footnote{This definition implies that when the body of an elementary random procedure is not a constant, its distribution function represents the marginal probability over any other random choices made in evaluating the body. This becomes important for implementing query.}. 

To provide an initial set of elementary random procedures we allow stochastic primitive functions, in addition to the usual constants, that randomly sample a return value depending only on the current environment.
Unlike other constants, these random functions are available only wrapped into elementary random procedures: {\tt (fun, args, env, dist)}, where $\text{\tt dist} = P( \text{ value } | \text{ {\tt env}, {\tt args}})$ is the probability function for {\tt fun}.  
We include several elementary random procedures, such as {\tt flip} which flips a fair coin (or flips a weighted coin when called with a weight argument).

A Church expression defines a generative process via the recursive evaluation procedure, {\tt eval}. This primitive procedure takes an expression and an environment and returns a value---it is an environment model, shared with Scheme, of Church's lambda calculus \cite{Church1932,r5rs}. The evaluation rules are given in Fig.~\ref{eval}.  %
An evaluation history for an expression {\tt $e$} is the sequence of recursive calls to {\tt eval}, and their return values, made by {\tt (eval '$e$ env)}. The probability of a finite evaluation history is the product of the probabilities for each elementary random procedure evaluation in this history\footnote{However, if evaluating an elementary random procedure results in evaluating another elementary random procedure we take only the probability of the first, since it already includes the second.}.  The \emph{weight} of an expression in a particular environment is the sum of the probabilities of all of its finite evaluation histories. An expression is \emph{admissible} in an environment if it has weight one, and a procedure is admissible if its body is admissible in its environment for all values of its arguments. An admissible expression defines a distribution on evaluation histories (we make this claim precise in section~\ref{formal}). Note that an admissible expression can have infinite histories, but the set of infinite histories must have probability zero. Thus admissibility can be thought of as the requirement that evaluation of an expression halts with probability one. Marginalizing this distribution over histories results in a distribution on values, which we write $\mu(\text{\tt e, env})$. 
Thus, {\tt (eval '$e$ env)}, for admissible {\tt $e$}, returns a sample from $\mu(\text{\tt e, env})$.

\begin{figure}
\begin{itemize}
\small
\itemsep 1pt
\item {\tt (eval '$c$ env)}: For constant $c$, return $c(\text{\tt env})$.
\item {\tt (eval '$x$ env)}: Look-up symbol $x$ in {\tt env}, return the value it is bound to.
\item {\tt (eval '($e_{1}$ $e_{2}$ ...)~env)}: Evaluate each {\tt (eval '$e_{i}$ env)}. The value of {\tt (eval '$e_{1}$ env)} should be a procedure ({\tt body}, {\tt $x_{2}$ ...}, {\tt env2}). Make {\tt env3} by extending {\tt env2}, binding {\tt $x_{2}$ ...} to the return values of  {\tt $e_{2}$ ...}. Return the value of {\tt (eval body env3)}.
\item {\tt (eval '(lambda ($x ...$)~$e$) env)}: Return the procedure ($e$, $x ...$, {\tt env}).
\item {\tt (eval '(if $e_{1}$ $e_{2}$ $e_{3}$) env)}: If {\tt (eval $e_{1}$ env)} returns {\tt True} return the return value of {\tt (eval $e_{2}$ env)}, otherwise of {\tt (eval $e_{3}$ env)}.
\item {\tt (eval '(quote $e$) env)}: Return the expression $e$ (as a value).
\item {\tt (eval '(define $x$ $e$) env)}: Extend {\tt env} by binding the value of {\tt (eval '$e$ env)} to $x$; return the extended environment.
\end{itemize}
\caption{An informal definition of the {\tt eval} procedure. If preconditions of these descriptions fail the constant value {\tt error} is returned. Note that constants represent  (possibly stochastic) functions from environments to values---truly ``constant'' constants return themselves.}
\label{eval}
\end{figure}

The procedure {\tt eval} allows us to interpret Church as a language for generative processes, but for useful probabilistic inference we must be able to sample from a distribution conditioned on some assertions (for instance the posterior probability of a hypothesis conditioned on observed data). The procedure {\tt (query '$e$ $p$ env)} is defined to be a procedure which samples a value from $\mu(\text{\tt e, env})$ conditioned on the predicate procedure $p$ returning {\tt True} when applied to the value of {\tt (eval '$e$ env)}. 
The environment argument {\tt env} is optional, defaulting to the current environment.
(Note that the special case of {\tt query} when the predicate $p$ is the constant procedure {\tt (lambda (x) True)} defines the same distribution on values as {\tt eval}.) For example, one might write {\tt (query '(pair (flip) (flip)) (lambda (v) (+ (first v) (last v))))} to describe the conditional distribution of two flips given that at least one flip landed heads.
If $e$ or $p$ are not admissible in {\tt env} the query result is undefined.
We describe this conditional distribution, and conditions for its well-definedness, more formally in Theorem \ref{bigwd}.
In Section \ref{montecarlo} we consider Monte Carlo techniques for implementing {\tt query}.

It can be awkward in practice to write programs using {\tt query}, because many random values must be explicitly passed from the query expression to the predicate through the return value. An alternative is to provide a means to name random values which are shared by all evaluations, building up a ``random world'' within the query.
To enable a this  style of programming,  we provide the procedure {\tt lex-query} (for ``lexicalizing query'') which has the form:
{\tt
\begin{tabbing}\small
(lex-query\\
\hspace{1em}\= '(\=(A {\it A-definition})\\
\> \>(B {\it B-definition})\\
\> \>    ...)\\
\> '$e$ '$p$)
\end{tabbing}}
where the first argument binds a lexicon of symbols to definitions, which are available in the environment in which the remaining (query and predicate) expressions are evaluated. In this form the predicate is an expression, and 
the final environment argument is omitted---the current environment is used.

A program in Church consists of a sequence of Church expressions---this sequence is called the \emph{top level}. Any definitions at the top level are treated as extending the global (i.e.~initial) environment, which then is used to evaluate the remaining top-level expressions. For instance:\\ 
{\tt \small (define A $e_{1}$) $e_{2}$}\\
is treated as:\\
{\tt \small (eval '$e_{2}$ (eval '(define A $e_{1}$) global-env))}.

\subsection{Stochastic Memoization}

In deterministic computation, memoization is a technique for efficient implementation that does not affect the language semantics: the first time a (purely functional) procedure is evaluated with given arguments its return value is recorded; thereafter evaluations of that procedure with those arguments directly return this value, without re-evaluating the procedure body. Memoization of a stochastic program can radically change the semantics: if {\tt flip} is an ordinary random procedure {\tt (= (flip) (flip))} is {\tt True} with probability $0.5$, but if {\tt flip} is memoized this expression is {\tt True} with probability one. More generally, a collection of memoized functions has a random-world semantics as discussed in \cite{McAllester2008}.
In Section \ref{examples} we use memoization together with {\tt lex-query} to describe generative processes involving an unknown number of objects with persistent features, similar to the BLOG language \cite{Milch2005}.

To formally define memoization in Church, we imagine extending the notion of environment to allow countably many variables to be bound in an environment. The higher-order procedure {\tt mem} takes an admissible-procedure and returns another procedure: if {\tt (eval $e$ env)} returns the admissible procedure {\tt (body, args, env2)}, then
{\tt (eval '(mem $e$) env)} returns the \emph{memoized procedure} {\tt (mfun$_{e}$, args, env+)}, where:
\begin{itemize}
\item {\tt env+} is {\tt env2} (notionally) extended with a symbol {\tt V$_{val}$}, for each value $val$, bound to a value drawn from the distribution $\mu(\text{\tt($e$ $val$), env})$.
\item {\tt mfun$_{e}$} is a new constant function such that {\tt mfun$_{e}$} applied to the environment {\tt env+} extended with {\tt args} bound to $val$ returns the value bound to {\tt V$_{val}$}.
\end{itemize}
This definition implies that infinitely many random choices may be made when a memoized random procedure is created---the notion of admissibility must be extended to expressions which involve {\tt mem}. In the next section we describe an appropriate extension of admissibility, such that admissible expressions still define a marginal distribution on values, and the conditional distributions defining {\tt query} are well-formed.

Ordinary memoization becomes a semantically meaningful construct within stochastic languages. This suggests that there may be useful generalizations of {\tt mem}, which are not apparent in non-stochastic computation. Indeed, instead of always returning the initial value or always re-evaluating, one could stochastically decide on each evaluation whether to use a previously computed value or evaluate anew. 
We define such a stochastic memoizer {\tt DPmem} by using the Dirichlet process (DP) \cite{Sethuraman1994}---a distribution on discrete distributions built from an underlying base measure. For an admissible procedure {\tt e}, the expression {\tt (DPmem  $a$ $e$)} evaluates in {\tt env} to a procedure which samples from a (fixed) sample from the DP with base measure $\mu(\text{\tt $e$, env})$ and concentration parameter $a$. (When $a{=}0$, {\tt DPmem} reduces to {\tt mem}, when $a{=}\infty$, it reduces to the identity.)
The notion of using the Dirichlet process to cache generative histories was first suggested in \citet{Johnson2007}, in the context of grammar learning.
In Fig.~\ref{fig:DP} we write the Dirichlet Process and {\tt DPmem} directly in Church, via a stick-breaking representation. This gives a definition of these objects, proves that they are semantically well-formed (provided the rest of the language is), and gives one possible implementation.
\begin{figure}
\hrule
\vspace{2pt}
\small
\begin{verbatim}
(define (DP alpha proc)
  (let ((sticks (mem (lambda x (beta 1.0 alpha))))
        (atoms  (mem (lambda x (proc)))))
    (lambda () (atoms (pick-a-stick sticks 1)))))

(define (pick-a-stick sticks J)
  (if (< (random) (sticks J))
      J
      (pick-a-stick sticks (+ J 1))))

(define (DPmem alpha proc)
  (let ((dps (mem (lambda args 
                    (DP alpha 
                        (lambda () (apply proc args)) 
                        )))))
    (lambda argsin ((apply dps argsin))) ))
\end{verbatim}
\hrule
\caption{Church implementation of the Dirichlet Process, via stick breaking, and {\tt DPmem}. (Evaluating {\tt (apply proc args)} in {\tt env} for {\tt args}={\tt (a$_{1}$ ...)} is equivalent to {\tt (eval '(proc a$_{1}$ ...)~env)}.)}
\label{fig:DP}
\end{figure}

We pause here to explain choices made in the language
definition. Programs written with pure functions, those that always
return the same value when applied to the same arguments, have a
number of advantages.  It is clear that a random function cannot be
pure, yet there should be an appropriate generalization of purity
which maintains some locality of information.  We believe the right
notion of purity in a stochastic language is \emph{exchangeability}:
if an expression is evaluated several times in the same environment,
the distribution on return values is invariant to the order of
evaluations.  This exchangeability is exploited by the
Metropolis-Hastings algorithm for approximating {\tt query} given in Section \ref{montecarlo}.

Mutable state (or an unpleasant, whole-program transformation into {\em continuation passing style}) is necessary to implement Church, both to model
randomness and to implement {\tt mem} using finite
computation. However, this statefulness preserves exchangeability.
Understanding the ways in which other stateful language constructs---in particular, primitives for the construction and modification of
mutable state---might aid in the description of stochastic processes remains
an important area for future work.

\subsection{Semantic Correctness}
\label{formal}
In this section we give formal statements of the claims above, needed to specify the semantics of Church, and sketch their proofs.
Let Church$^{-}$ denote the set of Church expressions that do not include {\tt mem}.
\begin{lem}
If $e \in \text{Church$^{-}$}$ then the weight of {\tt e} in a given environment is well-defined and $\leq 1$.
\label{wdnomem}
\end{lem}
\begin{proof}[Proof sketch]
Arrange the recursive calls to {\tt eval} into a tree with an evaluation at each node and edges connecting successive applications of {\tt eval}---if a node indicates the evaluation of an elementary random procedure there will be several edges descending from this node (one for each possible return value), and these edges are labeled with their probability. A history is a path from root to leaf in this tree and its probability is the product of the labels along the path. Let $W_{n}$ indicate the sum of probabilities of paths of length $n$ or less. The claim is now that $\lim_{n\rightarrow \infty} W_{n}$ converges and is bounded above by $1$. The bound follows because the sum of labels below any random node is $1$; convergence then follows from the monotone convergence theorem because the labels are non-negative.
\end{proof}

We next extend the notion of admissibility to arbitrary Church expressions involving {\tt mem}.
To compute the probability of an evaluation history we must include the probability of calls to {\tt mem}---that is, the probability of drawing each return value {\tt V$_{val}$}. Because there are infinitely many {\tt V$_{val}$}, the probability of many histories will then be zero, therefore we pass to equivalence classes of histories. Two histories are \emph{equivalent} if they are the same up to the values bound to {\tt V$_{val}$}---in particular they must evaluate all memoized procedures on the same arguments with the same return values. The probability of an equivalence class of histories is the marginal probability over all unused arguments and return values, and this is non-zero. The weight of an expression can now be defined as the sum over equivalence classes of finite histories.
\begin{lem}
The admissibility of a Church expression in a given environment is well defined, and any expression {\tt e} admissible in environment {\tt env} defines a distribution $\mu(\text{\tt e, env})$ on return values of {\tt (eval '$e$ env)}.
\label{wd}
\end{lem}
\begin{proof}[Proof sketch:]
The proof is by induction on the number of times {\tt mem} is used. Take as base case expressions without {\tt mem}; by Lemma~\ref{wdnomem} the weight is well defined, so the set of admissible expressions is also well defined. 

Now, assume $p = \text{\tt (body, args, env)}$ is an admissible procedure with well defined distribution on return values. The return from {\tt (mem p)} is well defined, because the underlying measure $\mu(\text{\tt p, env})$ is well defined.
It is then straightforward to show that any expression involving {\tt (mem p)}, but no other new memoized procedures, has a well defined weight. The induction step follows.

A subtlety in this argument comes if one wishes to express recursive memoized functions such as:\\
{\tt \small (define F (mem (lambda (x) (... F ...))))}. \\
Prima facie this recursion seems to eliminate the memoization-free base case. However,  any recursive definition (or set of definitions) may be re-written without recursion in terms of a fixed-point combinator: {\tt \small (define F (fix ...))}. With this replacement made we are reduced to the expected situation---application of {\tt fix} may fail to halt, in which case {\tt F} will be inadmissible, but the weight is well defined.
\end{proof}

\begin{figure}
\hrule
\vspace{2pt}
\small
\begin{comment} This function provides persistent class assignments to objects, where classes are symbols drawn from a pool with DP prior:\end{comment}
\begin{verbatim}
(define drawclass (DPmem 1.0 gensym))
(define class (mem (lambda (obj) (drawclass))))
\end{verbatim}
\begin{comment} For the beta-binomial model there's a coin weight for each feature/class pair, and each object has features that depend only on it's type: \end{comment}
\begin{verbatim}
(define coin-weight 
  (mem (lambda (feat obj-class) (beta 1 1))) )
(define value 
  (mem (lambda (obj feat)
         (flip (coin-weight feat (class obj))) )))
\end{verbatim}
 \begin{comment} For a gaussian-mixture on continuous data (with known variance), we just change the code for generating values: \end{comment}
 \begin{verbatim}
(define mean 
  (mem (lambda (obj-class) (normal 0.0 10.0))) )
(define cont-value 
  (mem (lambda (obj)
         (normal (mean (class obj)) 1.0) )))
\end{verbatim}
\begin{comment} The infinite relational model \cite{Kemp2006} with continuous data is similar, but means depend on classes of two objects: \end{comment}
\begin{verbatim}
(define irm-mean 
  (mem (lambda (obj-class1 obj-class2)
         (normal 0.0 10.0) )))
(define irm-value
  (mem (lambda (obj1 obj2)
         (normal (irm-mean (class obj1) (class obj2)) 
                 1.0 ))))
\end{verbatim}
\hrule
\caption{Church expressions for infinite mixture type models, showing use of the random-world programming style in which objects have persistent properties. Functions {\tt beta} and {\tt normal} generate samples from these standard distributions.}
\label{fig:crp-mix}
\end{figure}

Lemma \ref{wd} only applies to expressions involving {\tt mem} for admissible procedures---a relaxation is possible for partially admissible procedures in some situations. From Lemma~\ref{wd} it is straightforward to prove:
\begin{thm}
\label{bigwd}
Assume expression {\tt e} and procedure $p$ are admissible in {\tt env}, and let $V$ be a random value distributed according to $\mu({\tt e}, {\tt env})$.
If there exists a value $v$ in the support of $\mu({\tt e}, {\tt env})$ and {\tt True} has non-zero probability under $\mu(\text{\tt ($p$ $v$)}, {\tt env})$, then the conditional probability
\begin{equation*}
P(V {=} val \mid \text{\tt(eval '($p$ $V$) env)} {=} {\tt True})
\end{equation*}
is well defined.
\end{thm}
Theorem \ref{bigwd} shows that {\tt query} is a well-posed procedure; in Section \ref{montecarlo} we turn to the technical challenge of actually implementing query.

\section{EXAMPLE PROGRAMS}
\label{examples}

\begin{figure*}[t]
\hrule
\vspace{2pt}
\begin{tabular}{cc}
\begin{minipage}{\columnwidth}
\small
\begin{comment}
This deterministic higher-order function defines the basic structure of stochastic transition models:
\end{comment}
\begin{verbatim}
(define (unfold expander symbol)
  (if (terminal? symbol)
      symbol
      (map (lambda (x) (unfold expander x)) 
           (expander symbol) )))
\end{verbatim}
\begin{comment}
A Church model for a PCFG transitions via a fixed multinomial over expansions for each symbol:
\end{comment}
\begin{verbatim}
(define (PCFG-productions symbol)
  (cond ((eq? symbol 'S) 
         (multinomial '((S a) (T a)) '(0.2 0.8)) )
        ((eq? symbol 'T) 
         (multinomial '((T b) (a b)) '(0.3 0.7)) ))
        
(define (sample-pcfg) (unfold PCFG-productions 'S))
\end{verbatim}
\begin{comment}
The HDP-HMM \citep{Beal2002} uses memoized symbols for states and memoizes
transitions: 
\end{comment}
\begin{verbatim}
(define get-symbol (DPmem 1.0 gensym))    
(define get-observation-model 
  (mem (lambda (symbol) (make-100-sided-die))))
(define ihmm-transition 
  (DPmem 1.0 (lambda (state)  
               (if (flip) 'stop (get-symbol)) 
               
               
                   
\end{verbatim}
\end{minipage}
&
\begin{minipage}{\columnwidth}
\small
\begin{verbatim}                              
(define (ihmm-expander symbol)       
  (list ((get-observation-model symbol)) 
        (ihmm-transition symbol) )) 
(define (sample-ihmm) (unfold ihmm-expander 'S))    
\end{verbatim}
\begin{comment}The HDP-PCFG \citep{Liang2007} is also straightforward:\end{comment}
\begin{verbatim}
(define terms      '( a  b  c  d))
(define term-probs '(.1 .2 .2 .5))
(define rule-type
  (mem (lambda symbol)
         (if (flip) 'terminal 'binary-production))
(define ipcfg-expander 
  (DPmem 1.0 
         (lambda (symbol)
           (if (eq? (rule-type symbol) 'terminal)
               (multinomial terms term-probs)
               (list (get-symbol) (get-symbol))))
(define (sample-ipcfg) (unfold ipcfg-expander 'S))
\end{verbatim}
\begin{comment}
Making adapted versions of any of these models \citep{Johnson2007} only requires stochastically memoizing {\tt unfold}:
\end{comment}
\begin{verbatim}
(define adapted-unfold
  (DPmem 1.0 
         (lambda (expander symbol)
           (if (terminal? symbol)
                symbol
                (map (lambda (x) 
                       (adapted-unfold expander x)) 
                     (expander symbol) )))))
                     \end{verbatim}
\end{minipage}
\end{tabular}
\hrule
\caption{Some examples of ``stochastic transition models''.}
\label{fig:stochtrans}
\end{figure*}

In this section we describe a number of example programs, stressing the ability of Church to express a range of standard generative models. 
As our first example, we describe diagnostic causal reasoning in a simple scenario: given that the grass is wet on a given day, did it rain (or did the sprinkler come on)? 
In outline of this might take the form of the query:

{\small\begin{verbatim}
(lex-query
 '((grass-is-wet ...)
   (rain ...)
   (sprinkler ...)
 '(rain 'day2) 
 '(grass-is-wet 'day2) )\end{verbatim}}

where we define a causal model by defining functions that describe whether it rained, whether the sprinkler was on, and whether the grass is wet. The function {\tt grass-is-wet} will depend on both {\tt rain} and {\tt sprinkler}---first we define a noisy-or function:

{\small\begin{verbatim}
(define (noisy-or a astrength b bstrength baserate)
  (or (and (flip astrength) a)  
      (and (flip bstrength) b)
      (flip baserate)))\end{verbatim}} 
      
Using this noisy-or function, and a look-up table for various weights, we can fill in the causal model:

{\small\begin{verbatim}
(lex-query
 '((weight (lambda (ofwhat)
    (case ofwhat 
     (('rain-str) 0.9)
     (('rain-prior) 0.3)
      ..etc..)))
   (grass-is-wet (mem (lambda (day) 
     (noisy-or 
       (rain day) (weight 'rain-str)
       (sprinkler day) (weight 'sprinkler-str)
       (weight 'grass-baserate)))))
   (rain (mem (lambda (day) 
     (flip (weight 'rain-prior)))))
   (sprinkler (mem (lambda (day) 
     (flip (weight 'sprinkler-prior))))))
  '(rain 'day2) 
  '(grass-is-wet 'day2) )\end{verbatim}}
Note that we have used {\tt mem} to make the {\tt grass-is-wet}, {\tt rain}, and {\tt sprinkler} functions persistent. For example, {\tt (= (rain 'day2) (rain 'day2))} is always {\tt True} (it either rained on day two or not), this is necessary since both the query and predicate expressions will evaluate {\tt (rain 'day2)}. 

A Bayes net representation of this example would have clearly exposed the dependencies involved (though it would need to be supplemented with descriptions of the form of these dependencies). The Church representation, while more complex, lends itself to intuitive extensions that would be quite difficult in a Bayes net formulation.
For instance, what if we don't know the Bernoulli weights, but we do have observations of other days? We can capture this by drawing the weights from a hyper-prior, redefining the weight function to:

{\small\begin{verbatim}
...(weight (mem (lambda (ofwhat) (beta 1 1))))...\end{verbatim}}
If we now query conditioned on observations from other days, we implicitly learn the weight parameters of the model:
{\small\begin{verbatim}
(lex-query
  '...model definitions... 
  '(rain 'day2) 
  '(and 
    (grass-is-wet 'day1)
    (rain 'day1)
    (not (sprinkler 'day1))
    (grass-is-wet 'day2)) )\end{verbatim}}
Going further, perhaps the probability of rain depends on (unknown) types of days (e.g.~those with cumulus clouds, cirrus clouds, etc.), and perhaps the probability of the sprinkler activating depends on orthogonal types of days (e.g.~Mondays and Fridays versus other days). We can model this scenario by drawing the prior probabilities from two stochastically memoized beta distributions:

{\small\begin{verbatim}
(lex-query
 '((new-rain-prob 
     (DPmem 1.0 (lambda () (beta 1 1))))
   (new-sprinkler-prob 
     (DPmem 1.0 (lambda () (beta 1 1))))
   (rain (mem (lambda (day) 
     (flip (new-rain-prob)))))
   (sprinkler (mem (lambda (day) 
     (flip (new-sprinkler-prob))))))
  ...)\end{verbatim}}
  
With this simple change we have extended the original causal model into an infinite mixture of such models, in which days are co-clustered into two sets of types, based on their relationship to the wetness of the grass.

In the previous example we left the types of days implicit in the memoizer, using only the probability of rain or sprinkler. In Fig.~\ref{fig:crp-mix} we have given Church implementations for several infinite mixture models \cite[see][]{Kemp2006} using a different idiom---making the types into persistent properties of objects, drawn from an underlying memoized {\tt gensym} (recall that {\tt gensym} is simply a procedure which returns a unique value on each evaluation). Once we have defined the basic structure, {\tt class} to draw latent classes for objects, it is straightforward to define the latent information for each class (e.g.~{\tt coin-weight}), and the observation model (e.g.~{\tt value}). This basic structure may be used to easily describe more complicated mixture models, such as the continuous-data infinite relational model (IRM) from \cite{Kemp2006}. Fig.~\ref{fig:crp-mix} describes forward sampling for these models; to describe a conditional model, these definitions must be made within the scope of a query. For instance, if we wished to query whether two objects have the same class, conditioned on observed features:

{\small \begin{verbatim}
(lex-query
 '((drawclass (mem 1.0 gensym))
   (class ...)
   (coin-weight ...)
   (value ...))
 '(= (class 'alice) (class 'bob)) 
 '(and    
    (= (value 'alice 'blond) 1)
    (= (value 'bob 'blond) 1)
    (= (value 'jim 'blond) 0)))\end{verbatim}}

Another idiom (Fig.~\ref{fig:stochtrans}) allows us to write the common class of ``stochastic transition'' models, which includes probabilistic context free grammars (PCFGs), hidden Markov models (HMMs), and their ``infinite'' analogs. Writing the HDP-PCFG \cite{Liang2007} and HDP-HMM \cite{Beal2002} in Church provides a compact and clear specification to these complicated non-parametric models. If we memoize {\tt unfold} and use this {\tt adapted-unfold} on PCFG transitions we recover the Adaptor Grammar model of \cite{Johnson2007}; if we similarly ``adapt'' the HDP-PCFG or HDP-HMM we get interesting new models that have not been considered in the literature.

Fig.~\ref{fig:planning}(top) gives an outline for using Church to
represent planning problems. This is based on the translation of
planning into inference, given in \citet{Toussaint2006}, in which
rewards are transformed into the probability of getting a single
``ultimate reward''. Inference on this representation results in
decisions which soft-maximizes the expected
reward. Fig.~\ref{fig:planning}(bottom) fills in this framework for a
simple ``red-light'' game: the state is a light color (red/green) and
an integer position, a ``go'' action advances one position forward
except that going on a red light results in being sent back to
position 0 with probability {\tt cheat-det}. The goal is to be past
position 5 when the game ends; other rewards (e.g. for a staged game)
could be added by adding {\tt sp2}, {\tt sp3}, and so on.

\begin{figure}[tb]
\hrule
\vspace{2pt}
\small
\begin{verbatim}  
(define (transition state-action)
  (pair 
    (forward-model state-action)
    (action-prior) ))
(define (terminal? symbol) (flip gamma))
(define (reward-pred rewards) 
  (flip ((/ (sum rewards) (length rewards)))))
(lex-query 
  '((first-action (action-prior))
    (final-state 
      (first (unfold transition 
        (pair start-state first-action) )))
    (reward-list 
      (list (sp1 final-state) 
            (sp2 final-state) 
            ..etc.. ))
  'first-action
  '(reward-pred reward-list))
\end{verbatim}
\hrule
\begin{verbatim}
(define (forward-model s-a)
  (pair 
    (if (flip 0.5) 'red-light 'green-light)
    (let ((light (first (first s-a))) 
           (position (last (first s-a))) 
           (action (last s-a)))
      (if (eq? action 'go)
          (if (and (eq? light 'red-light) 
                   (flip cheat-det)) 
              0 
              (+ position 1))
          position))))
(define (action-prior) (if (flip 0.5) 'go 'stop))
(define (sp1 state) (if (> (last state) 5) 1 0))
\end{verbatim}
\hrule
\caption{Top: The skeleton of planning-as-inference in Church (inspired by \cite{Toussaint2006}). For simplicity, we assume an equal reward amount for each boolean ``state property'' that is true. Reward is given only when the state reaches a ``terminal state'', however the stochastic termination decision given by {\tt terminal?} results in an infinite horizon with discount factor {\tt gamma}. 
Bottom: A specific planning problem for the ``red-light'' game.}
\label{fig:planning}
\end{figure}

\section{CHURCH IMPLEMENTATION}
 \label{montecarlo}

Implementing Church involves two complications beyond the implementation of {\tt eval} as shown in Fig.~\ref{eval} (which is essentially the same as any lexically scoped, applicative order, pure Lisp \cite{r5rs}). First, we must find a way to implement {\tt mem}
 without requiring infinite structures (such as the {\tt V$_{val}$}). Second, we must implement {\tt query} by devising a means to sample from the appropriate conditional distribution.

To implement {\tt mem} we first note that the countably many {\tt V$_{val}$} are not all needed at once: they can be created as needed, extending the environment {\tt env+} when they are created. (Note that this implementation choices is stateful, but may be implemented easily in full Scheme: the argument/return value pairs can be stored in an association list which grows as need.)\footnote{A further optimization implements {\tt DPmem} via the Chinese restaurant process
representation of the DP \cite{Pitman2002}.}

\begin{SCfigure}
\begin{tabular}{c}
\includegraphics*[width=4.5cm, height=2.5cm, viewport=0 0 350 220]{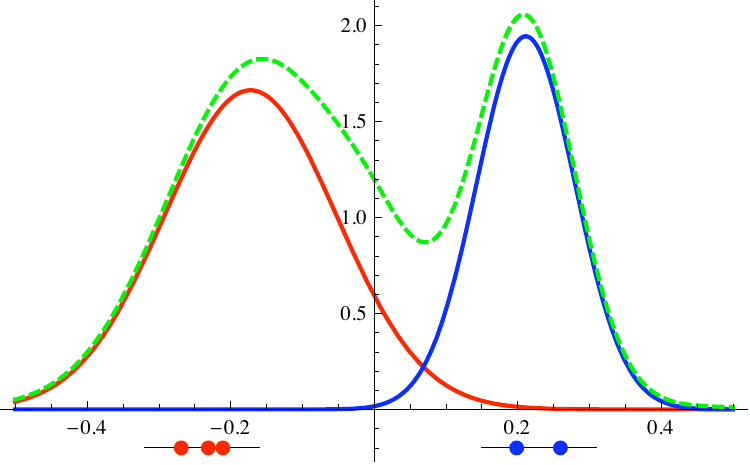}\\
\includegraphics*[width=4.5cm, height=2.5cm, viewport=0 0 350 220]{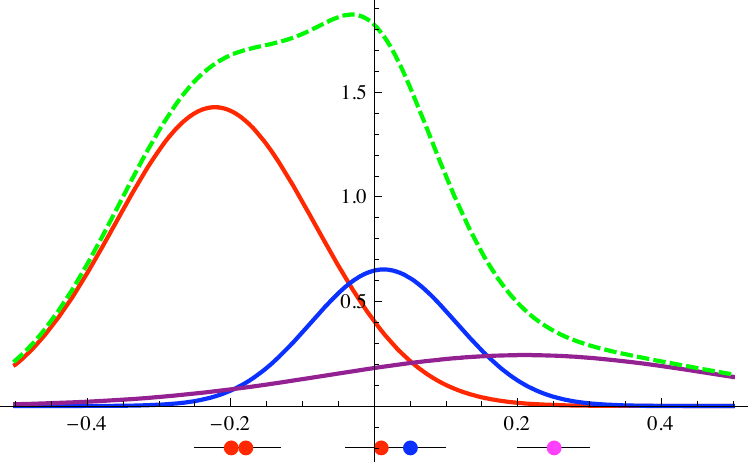}
\end{tabular}
\caption{Posterior samples from the infinite gaussian-mixture (with unknown variance) of Section \ref{examples}, using the collapsed rejection algorithm for {\tt query}. Two data-sets are shown (as dots) with mixture components and posterior predictive distribution.}
\label{lwresults}
\end{SCfigure}

We now turn to {\tt query}. The sampling-based semantics of Church allows us to define a simple rejection sampler from the conditional distribution defining {\tt query}; we may describe this as a Church expression: 

{\small\begin{verbatim}
(define (query exp pred env)
  (let ((val (eval exp env))
    (if (pred val)
        val
        (query exp pred env)))))
\end{verbatim}}

The ability to write {\tt query} as a Church program---a metacircular \cite{Abelson} implementation---provides
a compelling argument for Church's 
modeling power. However, exact sampling using this algorithm will often be intractable.
It is straightforward to implement 
a collapsed rejection
sampler that integrates out randomness in the predicate procedure
(accepting or rejecting a {\tt val} with probability equal to the
marginal probability that {\tt (p val)} is true). We show results in Fig.~\ref{lwresults}
of this exact sampler used to query the infinite gaussian-mixture model from Section \ref{examples}. 

\begin{SCfigure}
\includegraphics*[width=4.5cm, height=3.5cm, viewport=30 180 550 590]{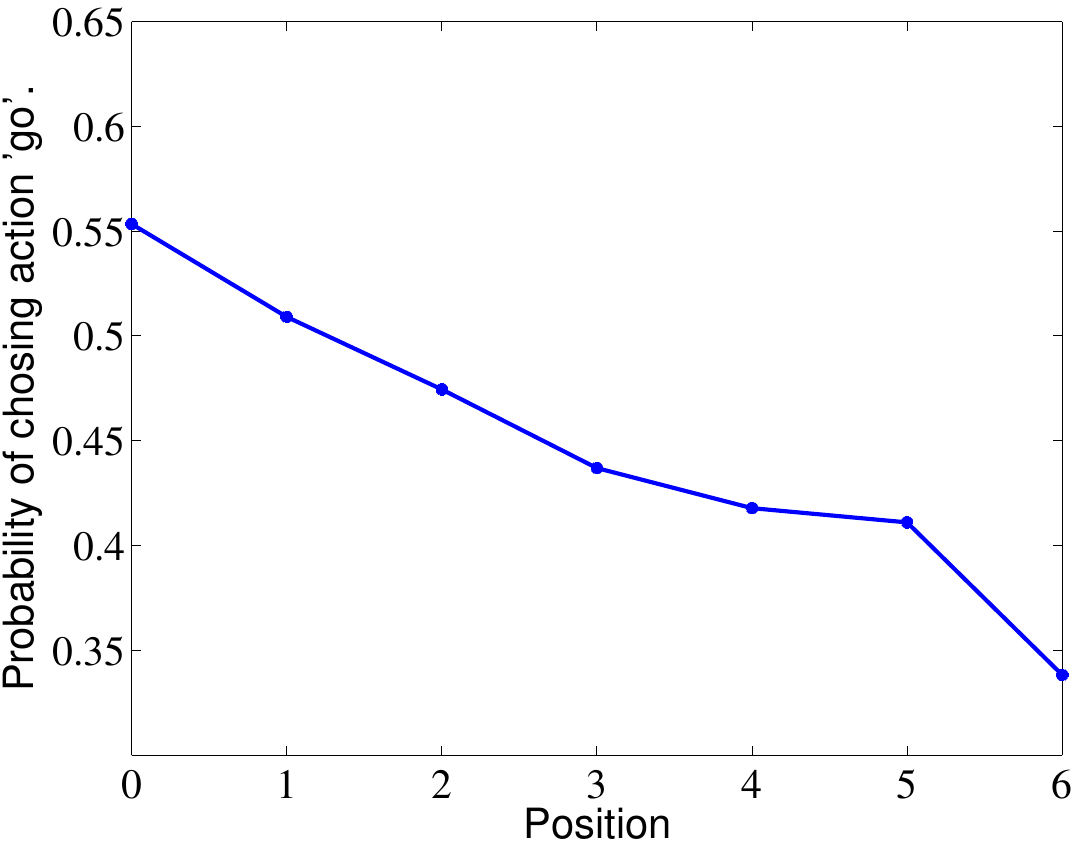}
\caption{Results from planning in the ``red-light'' game (Fig.~\ref{fig:planning}), showing the probability of ``cheating'' (going when the light is red) versus position. The goal is to end the game past position 5.}
\label{plan-res}
\end{SCfigure}

In Fig.~\ref{plan-res} we show the result of running the collapsed rejection query for planning in the ``red-light'' game, as shown in Fig.~\ref{fig:planning} (here {\tt gamma}${=}0.2$, {\tt cheat-det}${=}0.7$). The result is intuitive: when position is near 0 there is little to lose by ``cheating'', as position nears 5 (the goal line) there is more to loose, hence the probability of cheating decreases; once past the goal line there is nothing to be gained by going, so the probability of cheating drops sharply. Note that the ``soft-max'' formulation of planning used here results in fairly random behavior even in extreme positions.

\subsection{A Metropolis-Hastings Algorithm}

We now present a Markov chain Monte Carlo algorithm for approximately
implementing {\tt query}, as we expect (even collapsed) rejection sampling to
be intractable in general. Our algorithm executes stochastic local search over evaluation histories, making small changes by proposing changes to the return values of elementary random procedures.
These changes are constrained to produce the
conditioned result, collapsing out the predicate expression via its
marginal probability\footnote{Handling the rejection problem on chain
initialization (and queries across deterministic programs, more
generally) is a challenge. Replacing all language primitives (including
{\tt if}) with noisy alternatives and using tempering techniques provides one general solution, to be explored in future work.}. The use of evaluation histories, rather than values alone, can be viewed as an extreme form of data-augmentation: all random choices that lead to a value are made explicit in its history.

\begin{figure}
\includegraphics*[scale=0.6, viewport=-20 0 400 120]{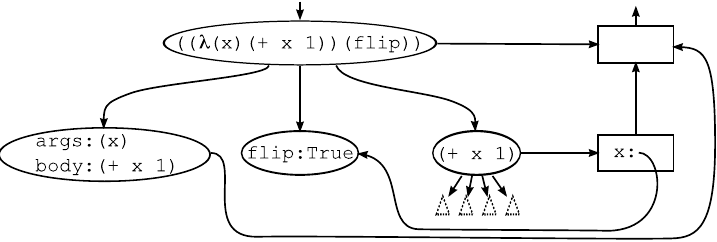}
\caption{A schematic computation trace.}
\label{trace}
\end{figure}

The key abstraction we use for MCMC is the {\em computation trace}.  A
computation trace
is a directed, acyclic graph composed of two connected trees.  The
first is a tree of evaluations, where an evaluation node points to
evaluation nodes for its recursive calls to {\tt eval}.  The second is
a tree of environment extensions, where the node for an extended
environment points to the node of the environment it extends.  The
evaluation node for each {\tt (eval '$e$ env)} points to the environment
node for {\tt env}, and evaluation nodes producing values to be
bound are pointed to by the environment extension of the binding.
Traces are in one-to-one correspondence with equivalence classes of evaluation
histories, described earlier\footnote{Also note that the acyclicity of traces is a direct result
of the purity of the Church language: if a symbol's value were
mutated, its environment would point to the evaluation node that
determined its new value, but that node would have been evaluated in
the same environment.}.
Fig.~\ref{trace} shows the fragment of a computation trace for
evaluation of the expression {\tt ((lambda (x) (+ x 3))
(flip))}. 

For each elementary random procedure {\tt p} we need a Markov chain transition kernel $K_{\text{\tt p}}$ that proposes a
new return value for that procedure given its current arguments. A generic such kernel comes from re-evaluating {\tt (eval '($p$ args) env)}; however, a proper Church standard library could
frequently supply more efficient proposal kernels for particular procedures (for instance a drift kernel for {\tt normal}). Our requirement is
that we are able to sample a proposal from $K_{\text{\tt p}}$ as well as evaluate its
transition probability $q_{\text{\tt p}}(\cdot|\cdot)$.

If we simply apply $K_{\text{\tt p}}$ to a trace, the trace can become ``inconsistent''---no longer representing a valid evaluation history from {\tt eval}. To construct a complete Metropolis-Hastings proposal from $K_{\text{\tt p}}$, we must keep the computation trace consistent,
and modify the proposal probabilities accordingly, by recursing along
the trace updating values and potentially triggering new evaluations.
For example, if we change the value of {\tt flip} in {\tt (if
(flip) $e_{1}$ $e_{2}$)} from {\tt False} to {\tt True} we must: absorb the probability of 
{\tt (eval $e_{2}$ env)} in the reverse proposal probability, evaluate {\tt
$e_{1}$} and attach it to the trace, and include the probability of the
resulting sub-trace in the forward proposal probability.
(For a particular trace, the probability of the sub-trace for expression {\tt $e$} is the probability of the equivalence class of evaluation histories corresponding to this sub-trace.)
 The recursions for trace consistency and proposal computation are delicate but
straightforward, and we omit the details due to space constraints\footnote{We implemented our MCMC algorithm atop the Blaise system \cite{Bonawitz}, which simplifies these recursively triggered
kernel compositions.}. Each step of our MCMC algorithm\footnote{At the time of writing we have not implemented this algorithm for programs that use {\tt mem}, though we believe the necessary additions to be straightforward.} consists of applying a kernel $K_{\text{\tt p}}$ to the evaluations of a randomly chosen elementary
random primitive in the trace, updating the trace to maintain consistency (collecting appropriate corrections to the proposal probability), and applying the Metropolis-Hastings criterion to accept or reject this proposal.
(This algorithm ignores some details needed for
queries containing nested queries, though we believe these to be straightforward.)

\begin{SCfigure}
\begin{tabular}{c}
\includegraphics*[width=4.5cm, height=3.3cm, viewport=65 200 555 600]{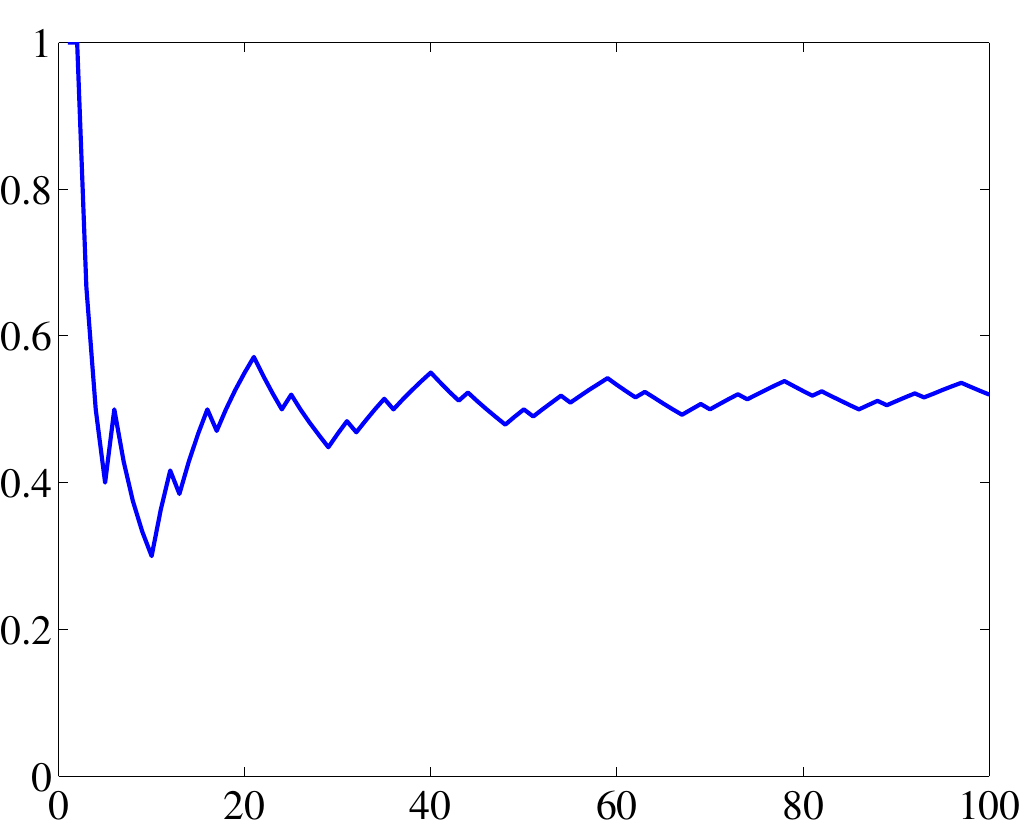}\\
\includegraphics*[width=4.5cm, height=3.3cm, viewport=65 180 555 600]{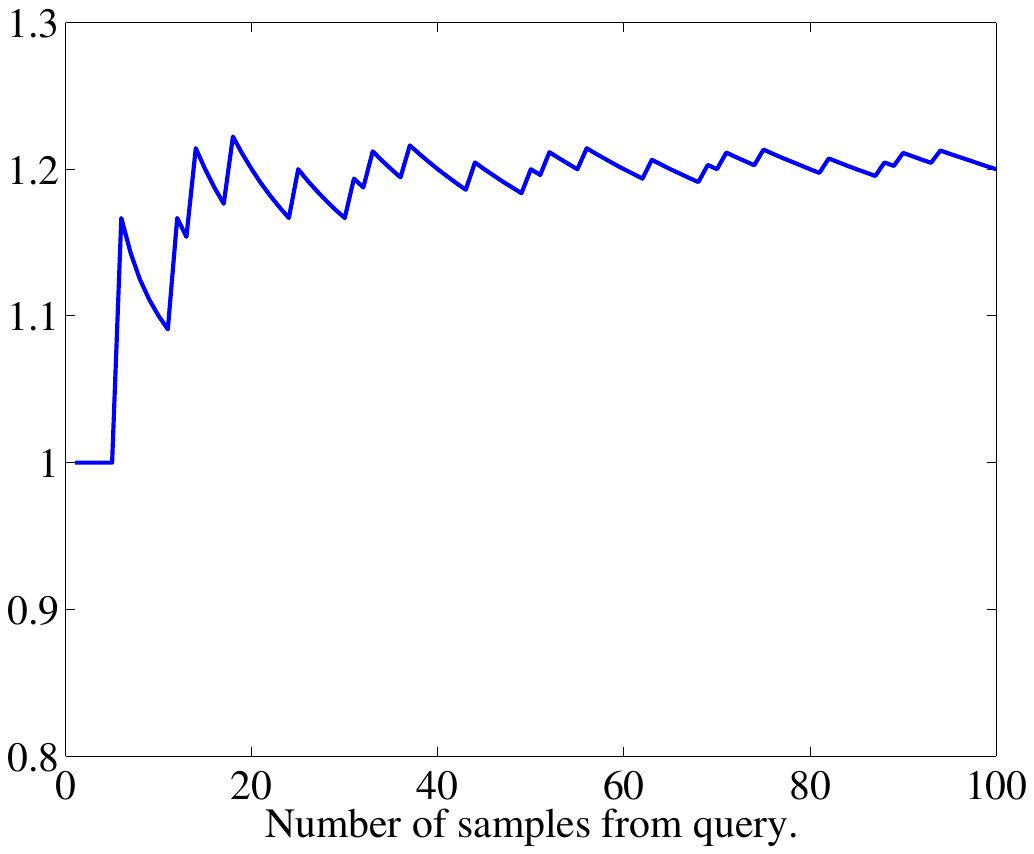}\\
\end{tabular}
\caption{Convergence of one run of the MCMC algorithm on the ``sprinkler'' example. 
(Each sample from query uses 30 MCMC steps.)
Top: The probability of {\tt rain}.
Bottom: The expected value of {\tt (+ (rain) (sprinkler))}, showing explaining away. The sum is slightly above $1.0$ because one cause is usually present, but both rarely are.}
\label{mcmcresults}
\end{SCfigure}

We have implemented and verified this algorithm on several examples that exercise all of the recursion and update logic of the system. In Fig.~\ref{mcmcresults} we have shown convergence results for this algorithm running on the simple ``sprinkler'' example of Section \ref{examples}.

\section{DISCUSSION}
\label{conclusion}

While Church builds on many other attempts to marry probability theory with computation, it is distinct in several important ways. First, Church is founded on the lambda calculus, allowing it to
represent higher-order logic and separating it from many
related languages. For example, unlike several widely used languages
grounded in propositional logic (e.g.~BUGS \cite{Lunn2000}) and first-order
logic (e.g.~the logic programming approaches of
\cite{Muggleton1996,Sato1997}, BLOG \cite{Milch2005}, and Markov logic \cite{Richardson2006}),
generative processes in Church are first-class objects that can be arbitrarily
composed and abstracted. The example programs in Section
\ref{examples} illustrate the representational flexibility of Church;
while some of these programs may be naturally represented in one or
another existing language, we believe that no other language can
easily represent all of these examples. 

The stochastic functional language IBAL \cite{Pfeffer2001}, based
on the functional language ML, is quite similar to Church, but the two
languages emphasize different aspects of functional
programming. Other related work includes non-determistic \cite{McCarthy1963} and weighted non-deterministic   \cite{Radul2007} extensions to Lisp. Unlike these approaches, the semantics of Church is fundamentally
sampling-based: the denotation of admissible expressions as
distributions follows from the semantics of evaluation rather than
defining it. 
This semantics, combined with dynamic typing (cf.~static typing of ML), permits the definition and exact implementation of
{\tt query}
as an ordinary
Church procedure, rather than a special transformation applied to the
distribution denoted by a program. Because {\tt query} is defined via sampling, describing
approximate inference is particularly natural within Church.

A number of the more unusual features of Church as a stochastic programming language derive from its basis in Lisp. Since {\tt query} and {\tt eval} are the basic
constructs defining the meaning of Church expressions, we have a
metacircular \cite{Reynolds1972} description of Church within Church. This provides
clarity in reasoning about the language, and allows self-reflection
within programs: queries may be nested within queries, and programs
may reason about programs. 
Church expressions can serve both as a declarative notation for uncertain beliefs
(via the distributions they represent) and as a
procedural notation for stochastic and deterministic processes (via evaluation).
Because expressions are themselves values, this generalizes the
Lisp unification of programs and data to a unification of stochastic
processes, Church expressions, and uncertain beliefs. These observations suggest
exciting new modeling paradigms. 
For instance, {\tt eval} nested
within {\tt query} may be used to learn programs, where the prior on
programs is represented by another Church program. Issues of
programming style then become issues of description length and
inductive bias. As another example, {\tt query} nested within {\tt query} may be used to represent an agent reasoning about another agent.

Of course, Church's representational flexibility comes at the cost of
substantially increased inference complexity. Providing efficient
implementations of {\tt query} is a critical challenge as our current implementation is not yet efficient enough
for typical machine learning applications; this may be
greatly aided by building on techniques used for inference in other
probabilistic languages
\cite[e.g.][]{McAllester2008,Pfeffer2001,Milch2005}. For example, in
Church, exact inference by enumeration could be seen as a program
analysis that transforms expressions involving {\tt query} into
expressions involving only {\tt eval}; identifying and exploiting
opportunities for such transformations seems appealing.

Probabilistic models and stochastic algorithms are finding
increasingly widespread use throughout artificial intelligence and
cognitive science, central to areas as diverse as vision,
planning, and natural language understanding. As their usage grows and
becomes more intricate, so does the need for formal languages
supporting model exchange, reuse, and machine execution. We hope
Church represents a significant step toward this goal. 

\subsubsection*{Acknowledgements} 
The authors would like to thank Gerry Sussman, Hal Abelson, Tom
Knight, Brian Milch, David McAllester and Alexey Radul for helpful discussions. This
work was funded in part by a grant from NTT Communication Sciences
Laboratory.

\bibliographystyle{plainnat}

\bibsep=2pt
\small
\bibliography{church}

\end{document}